\documentclass[a4paper,11pt]{article}
\usepackage{pos}
\usepackage{booktabs}

\title{Flavor-deconstructed neutrinos}

\author*[a]{Avelino Vicente}

\affiliation[a]{Instituto de F\'{i}sica Corpuscular, \\
  CSIC-Universitat de Val\`{e}ncia, 46980 Paterna, Spain}

\emailAdd{avelino.vicente@ific.uv.es}

\abstract{
A tentative approach to explain the flavor puzzle consists of embedding the Standard Model in a larger gauge symmetry that contains a separate gauge group for each fermion family. In such gauge non-universal (or flavor-deconstructed) theories, neutrinos pose some challenges. I will discuss existing ideas in the literature and present a simple model in which flavor deconstruction naturally leads to sequential dominance for both neutrinos and charged leptons, thus providing a viable explanation for the flavor structure of the lepton sector.
}

\FullConference{Proceedings of the Corfu Summer Institute 2025 "School and Workshops on Elementary Particle Physics and Gravity" (CORFU2025)\\
27 April - 28 September, 2025\\
Corfu, Greece\\}

\begin{document}
\maketitle

\section{Introduction}
\label{sec:intro}

The Standard Model (SM) of particle physics has been enormously successful in describing data from a wide variety of experimental sources. However, it fails to provide a fundamental understanding of several issues. A major example is the structure underlying the observed fermion masses and mixings. The SM can accommodate all measurements, but cannot explain why the masses of each type of charged fermion (up quarks, down quarks and charged leptons) are highly hierarchical, or why the mixing patterns differ so much between the quark and lepton sectors. These open questions are usually referred to as \textit{the flavor puzzle}.

Many theoretical ideas have been proposed over the years. Broadly speaking, they can be classified into two main categories. The traditional approach is based on \textbf{flavor symmetries}, where new horizontal symmetries are introduced in order to generate specific flavor patterns after being properly broken. Numerous examples exist in the literature, employing symmetries of various types: continuous or discrete, global or gauge.

An alternative approach to the flavor puzzle has gained popularity in recent years: \textbf{flavor deconstruction}. In flavor-deconstructed (or gauge non-universal) models, the SM is embedded in a larger gauge symmetry with a separate factor for each fermion family. This framework has proven very successful in explaining the hierarchies in charged fermion masses and the nearly diagonal structure of the quark mixing matrix. However, some challenges have been identified when addressing the lepton sector~\cite{Greljo:2024ovt}, mainly due to the apparently anarchical structure of the lepton mixing matrix. Below, we review this problem and discuss the solution proposed in~\cite{FernandezNavarro:2025zmb}. This proposal goes beyond the tri-hypercharge theory~\cite{FernandezNavarro:2023rhv} by further decomposing the three family hypercharges in a way that naturally reproduces the observed patterns in the lepton sector.

The rest of the paper is organized as follows: In Sec.~\ref{sec:deconstruction}, we introduce the general idea of flavor deconstruction and discuss the challenges it faces when applied to the lepton sector. Our proposal to address these issues, along with some of its predictions, is presented in Sec.~\ref{sec:proposal}. We conclude in Sec.~\ref{sec:final}.

\section{Flavor deconstruction}
\label{sec:deconstruction}

In flavor-deconstructed models, the SM gauge symmetry is replaced by
\begin{equation}
  G = G_{\mathrm{universal}}\times G_{1}\times G_{2}\times G_{3} \, ,
\end{equation}
where $G_{\mathrm{universal}}$ is a family-universal group (fermions belonging to the three families are charged under it), while each $G_i$ ($i=1,2,3$) is a separate factor associated with each fermion family. In addition, the SM Higgs doublet is a third-family field: a singlet under $G_1$ and $G_2$, but not under $G_3$. This implies that only the third fermion family has Yukawa interactions (and thus receives masses) at the renormalizable level,
\begin{equation}
  \mathcal{L} = y_t \, q_3 H u_3^c + y_b \, q_3 \widetilde{H} d_3^c + y_\tau \, \ell_3 \widetilde{H} e_3^c \, .
\end{equation}
This is an excellent starting point to address the flavor puzzle, as it already explains why the third family is heavier than the other two. The masses for the other two families, as well as the mixing structure, are induced by means of non-renormalizable operators with $\mathcal{O}(1)$ coefficients involving scalars with non-trivial representations under the flavor non-universal gauge factor. These scalars are responsible for the spontaneous breaking
\begin{equation}
  G_1 \times G_2 \times G_3 \to G_{1+2+3} \, ,
\end{equation}
where $G_{1+2+3}$ is identified with the flavor-universal SM group that has been deconstructed. This general setup successfully leads to a dynamical generation of the observed flavor pattern of the SM.

Flavor deconstruction has been relatively popular in the flavor community in recent years. For instance, models based on the deconstruction of the Pati–Salam group~\cite{Bordone:2017bld}, of $SU(2)_L$~\cite{Davighi:2023xqn} or of $U(1)_Y$ have been constructed, to name a few examples. The latter case goes under the name of \textit{tri-hypercharge}~\cite{FernandezNavarro:2023rhv} (see also~\cite{FernandezNavarro:2023hrf,FernandezNavarro:2024hnv,FernandezNavarro:2025zmb} for other works on complete hypercharge deconstruction and~\cite{Davighi:2023evx} for the case of partial deconstruction). It is given by $G_{\mathrm{universal}} = SU(3)_c \times SU(2)_L$ and $G_i = U(1)_{Y_i}$ and can be regarded as one of the simplest realizations of flavor deconstruction.

Flavor-deconstructed models naturally explain hierarchies. They typically generate large mass hierarchies for quarks and charged leptons, along with small quark mixing. However, when a neutrino mass mechanism is incorporated, these models also tend to produce small lepton mixing, in clear contrast with neutrino oscillation data~\cite{deSalas:2020pgw}. This issue has been discussed in several works, most prominently in~\cite{Greljo:2024ovt}.

\begin{table}
\centering %
\begin{tabular}{lcccc}
\toprule 
Field  & $U(1)_{Y_{1}}$  & $U(1)_{Y_{2}}$  & $U(1)_{Y_{3}}$  & $SU(3)_{c}\times SU(2)_{L}$\tabularnewline
\midrule 
$\ell_{1}$  & $-\frac{1}{2}$  & 0  & 0  & ($\mathbf{1},\mathbf{2}$)\tabularnewline
$\ell_{2}$  & 0  & $-\frac{1}{2}$  & 0  & ($\mathbf{1},\mathbf{2}$)\tabularnewline
$\ell_{3}$  & 0  & 0  & $-\frac{1}{2}$  & ($\mathbf{1},\mathbf{2}$)\tabularnewline
\midrule 
$e_{1}^{c}$  & 1 & 0  & 0  & ($\mathbf{1},\mathbf{1}$)\tabularnewline
$e_{2}^{c}$  & 0  & 1  & 0  & ($\mathbf{1},\mathbf{1}$)\tabularnewline
$e_{3}^{c}$  & 0  & 0  & 1 & ($\mathbf{1},\mathbf{1}$)\tabularnewline
\midrule 
$\nu_{2}^{c}$  & 0  & 0  & 0  & ($\mathbf{1},\mathbf{1}$)\tabularnewline
$\nu_{3}^{c}$  & 0  & 0  & 0  & ($\mathbf{1},\mathbf{1}$)\tabularnewline
\midrule 
$H_{u}$  & 0  & 0  & $\frac{1}{2}$  & ($\mathbf{1},\mathbf{2}$)\tabularnewline
$H_{d}$  & 0  & 0  & $-\frac{1}{2}$  & ($\mathbf{1},\mathbf{2}$)\tabularnewline
\midrule 
$\phi_{12}$  & $\frac{1}{2}$  & $-\frac{1}{2}$  & 0  & ($\mathbf{1},\mathbf{1}$)\tabularnewline
$\phi_{23}$  & 0  & $\frac{1}{2}$  & $-\frac{1}{2}$  & ($\mathbf{1},\mathbf{1}$)\tabularnewline
\bottomrule
\end{tabular}\caption{Minimal tri-hypercharge model for the lepton sector~\cite{FernandezNavarro:2024hnv}.
\label{tab:min-TH}}
\end{table}

To illustrate the challenge, let us consider the example of tri-hypercharge. In particular, we focus on the minimal tri-hypercharge model for the lepton sector introduced in~\cite{FernandezNavarro:2024hnv}. The SM particle content is extended by: (i) two Higgs doublets, $H_u$ and $H_d$ (instead of just one), to address the hierarchies between up- and down-type fermions; (ii) two right-handed neutrinos, $\nu_{2,3}^c$; and (iii) the $\phi_{12}$ and $\phi_{23}$ \textit{hyperons}. The latter are scalars with non-zero individual hypercharges but vanishing total hypercharge. When they acquire vacuum expectation values (VEVs), they break $U(1)_{Y_1} \times U(1)_{Y_2} \times U(1)_{Y_3} \to U(1)_Y$, where $Y = Y_1 + Y_2 + Y_3$ is the SM universal hypercharge. The minimal lepton sector, along with the corresponding charges under the gauge groups of the model, is shown in Table~\ref{tab:min-TH}.

With these representations, the neutrino Yukawa Lagrangian is given by
\begin{equation}
  \mathcal{L}=a_{3i}^{\nu} \ell_{3}H_{u}\nu_{i}^{c}+a_{2i}^{\nu}\frac{\phi_{23}}{\Lambda^{\nu}_{23}}\ell_{2}H_{u}\nu_{i}^{c}+a_{1i}^{\nu}\frac{\phi_{12}}{\Lambda^{\nu}_{12}}\frac{\phi_{23}}{\Lambda^{\nu}_{23}}\ell_{1}H_{u}\nu_{i}^{c}+M_{ij}\nu_{i}^{c}\nu_{j}^{c}+\mathrm{h.c.} \, ,
\end{equation}
with $a_{ij}^{\nu,e}$ some dimensionless coefficients. When the hyperons get their VEVs, they generate Dirac and Majorana mass matrices for the neutrinos,
\begin{equation}
  m_D = \left(\begin{array}{cc}
a_{11}^{\nu}\epsilon_{12}^{\nu}\epsilon_{23}^{\nu} & a_{12}^{\nu}\epsilon_{12}^{\nu}\epsilon_{23}^{\nu}\\
a_{21}^{\nu}\epsilon_{23}^{\nu} & a_{22}^{\nu}\epsilon_{23}^{\nu}\\
a_{31}^{\nu} & a_{32}^{\nu}
\end{array}\right) \, , \quad M_M = \left(\begin{array}{cc}
M_{22} & M_{23}\\
M_{32} & M_{33}
  \end{array}\right) \, ,
\end{equation}
as well as a Dirac mass matrix for the charged leptons,
\begin{equation}
  m_e = \left(\begin{array}{ccc}
a_{11}^{e}\epsilon_{12}^{e}\epsilon_{23}^{e} & a_{12}^{e}\epsilon_{12}^{e}\epsilon_{23}^{e} & a_{13}^{e}\epsilon_{12}^{e}\epsilon_{23}^{e}\\
a_{21}^{e}(\epsilon_{12}^{e}){}^{2}\epsilon_{23}^{e} & a_{22}^{e}\epsilon_{23}^{e} & a_{23}^{e}\epsilon_{23}^{e}\\
a_{31}^{e}(\epsilon_{12}^{e}){}^{2}(\epsilon_{23}^{e}){}^{2} & a_{32}^{e}(\epsilon_{23}^{e}){}^{2} & a_{33}^{e} \end{array}\right) \, .
\end{equation}
Here $\epsilon_{12}^{\nu}=\langle\phi_{12}\rangle/\Lambda_{12}^{\nu}$, $\epsilon_{23}^{\nu}=\langle\phi_{23}\rangle/\Lambda_{23}^{\nu}$, $\epsilon^{e}_{12}=\langle\phi_{12}\rangle/\Lambda^{e}_{12}$ and $\epsilon^{e}_{23}=\langle\phi_{23}\rangle/\Lambda^{e}_{23}$. All $\epsilon_{ij}^{\nu,e}$ are assumed to be small parameters due to $\Lambda_{ij}^{\nu,e} \gg \langle \phi_{ij} \rangle$. We see that $m_D$ is a very hierarchical matrix, $M_M$ is anarchic and $m_e$ is nearly diagonal and hierarchical. It is straightforward to check that with these ingredients, the resulting lepton mixing angles are small, unless the dimensionless $\mathcal{O}(1)$ coefficients $a_{ij}^\nu$ and $a_{ij}^e$ are tuned.

Several ways out to this problem have been proposed:
\begin{itemize}
\item Introduce extra scalars (e.g. hyperons) which only participate in the neutrino sector and change the Yukawa texture~\cite{FernandezNavarro:2023rhv}.
\item Go beyond the validity of the effective approach and generate lepton mixing in the full ultraviolet theory~\cite{FernandezNavarro:2024hnv}.
\item Consider particular gauge symmetries where both hierarchical $m_D$ and $m_M$ are hierarchical, but the overall hierarchies cancel in the neutrino mass matrix~\cite{Greljo:2024ovt}.
\item Charge all lepton doublets under the same factor (e.g. hypercharge) and assume that the resulting gauge anomalies are cancelled due to the presence of extra fermions at high energies~\cite{Fuentes-Martin:2024fpx,Lizana:2024jby}.
\end{itemize}

However, all these approaches lead to an anarchic effective neutrino mass matrix. All its entries are given in terms of $\mathcal{O}(1)$ coefficients which are fitted to neutrino oscillation data. In the next Section we will show an alternative solution leading to a mechanism beyond anarchy.

\section{A new proposal for the lepton sector}
\label{sec:proposal}

Our new proposal is based on the following key observation: in the example shown in the previous Section, the two right-handed neutrinos are both singlets under the tri-hypercharge gauge group, and hence are indistinguishable, which results in the two columns of the Dirac mass matrix being approximately equal and the Majorana matrix being anarchical. Therefore, the solution seems clear: extend tri-hypercharge to a larger gauge group under which the right-handed neutrinos are not singlets.

\begin{table}
\centering{}%
\begin{tabular}{lcccc}
\toprule 
Field  & $U(1)_{Y_{1}}$  & $U(1)_{R_{2}}\times U(1)_{(B-L)_{2}/2}$  & $U(1)_{R_{3}}\times U(1)_{(B-L)_{3}/2}$ & $SU(3)_{c}\times SU(2)_{L}$\tabularnewline
\midrule 
$\ell_{1}$  & $-\frac{1}{2}$  & $\mathrm{(0,0)}$  & $\mathrm{(0,0)}$ & ($\mathbf{1},\mathbf{2}$)\tabularnewline
$\ell_{2}$  & 0  & $\mathrm{(0,-\frac{1}{2})}$  & $\mathrm{(0,0)}$ & ($\mathbf{1},\mathbf{2}$)\tabularnewline
$\ell_{3}$  & 0  & $\mathrm{(0,0)}$  & $\mathrm{(0,-\frac{1}{2})}$ & ($\mathbf{1},\mathbf{2}$)\tabularnewline
\midrule 
$e_{1}^{c}$  & 1  & $\mathrm{(0,0)}$  & $\mathrm{(0,0)}$ & ($\mathbf{1},\mathbf{1}$)\tabularnewline
$e_{2}^{c}$  & 0  & $\mathrm{(\frac{1}{2},\frac{1}{2})}$  & $\mathrm{(0,0)}$ & ($\mathbf{1},\mathbf{1}$)\tabularnewline
$e_{3}^{c}$  & 0  & $\mathrm{(0,0)}$  & $\mathrm{(\frac{1}{2},\frac{1}{2})}$ & ($\mathbf{1},\mathbf{1}$)\tabularnewline
\midrule 
$\nu_{2}^{c}$  & 0  & $\mathrm{(-\frac{1}{2},\frac{1}{2})}$  & $\mathrm{(0,0)}$ & ($\mathbf{1},\mathbf{1}$)\tabularnewline
$\nu_{3}^{c}$  & 0  & $\mathrm{(0,0)}$  & $\mathrm{(-\frac{1}{2},\frac{1}{2})}$ & ($\mathbf{1},\mathbf{1}$)\tabularnewline
\midrule 
$H_{u,d}$  & 0  & $\mathrm{(0,0)}$  & $\mathrm{(\pm\frac{1}{2},0)}$ & ($\mathbf{1},\mathbf{2}$)\tabularnewline
\midrule 
$\chi_{2}$  & 0  & $\mathrm{(1,-1)}$  & $\mathrm{(0,0)}$ & ($\mathbf{1},\mathbf{1}$)\tabularnewline
$\chi_{3}$  & 0  & $\mathrm{(0,0)}$  & $\mathrm{(1,-1)}$ & ($\mathbf{1},\mathbf{1}$)\tabularnewline
\midrule 
$\phi_{12}^{R}$  & $\frac{1}{2}$  & $\mathrm{(-\frac{1}{2},0)}$  & $\mathrm{(0,0)}$ & ($\mathbf{1},\mathbf{1}$)\tabularnewline
$\phi_{12}^{L}$  & $\frac{1}{2}$  & $\mathrm{(0,-\frac{1}{2})}$  & $\mathrm{(0,0)}$ & ($\mathbf{1},\mathbf{1}$)\tabularnewline
$\phi_{23}^{R}$  & 0  & $\mathrm{(\frac{1}{2},0)}$  & $\mathrm{(-\frac{1}{2},0)}$ & ($\mathbf{1},\mathbf{1}$)\tabularnewline
$\phi_{23}^{L}$  & 0  & $\mathrm{(0,\frac{1}{2})}$  & $\mathrm{(0,-\frac{1}{2})}$ & ($\mathbf{1},\mathbf{1}$)\tabularnewline
\bottomrule
\end{tabular}\caption{Field content of the lepton sector.
  \label{tab:proposal}}
\end{table}

Therefore, we consider the gauge group
\begin{equation}
G = SU(3)_c \times SU(2)_L \times U(1)_{Y_{1}}\times U(1)_{R_{2}}\times U(1)_{(B-L)_{2}/2}\times U(1)_{R_{3}}\times U(1)_{(B-L)_{3}/2} \, .
\end{equation}
Hypercharge has been decomposed into $U(1)_R \times U(1)_{(B-L)/2}$ for the second and third families. We also note that $SU(3)_c$ and $SU(2)_L$ remain universal. The particle content of the new lepton sector is shown in Table~\ref{tab:proposal}. We have introduced the scalars $\chi_i$, which break $U(1)_{R_i} \times U(1)_{(B-L)_i/2} \to U(1)_{Y_i}$, with $i=2,3$, as well as new hyperons for the $U(1)_R$ and $U(1)_{(B-L)/2}$ groups. With these ingredients, and following a procedure analogous to that in the previous section, one obtains the following mass matrices:
\begin{equation}
  m_D = \left(\begin{array}{cc}
a_{12}^{\nu}\epsilon_{12}^{L}\epsilon_{23}^{R} & a_{13}^{\nu}\epsilon_{12}^{L}\epsilon_{23}^{L}\\
a_{22}^{\nu}\epsilon_{23}^{R} & a_{23}^{\nu}\epsilon_{23}^{L}\\
a_{32}^{\nu}\epsilon_{23}^{L}\epsilon_{23}^{R} & a_{33}^{\nu}
\end{array}\right) \, , \quad M_M = \left(\begin{array}{cc}
\langle \chi_{2} \rangle & \epsilon_{23}^{L}\epsilon_{23}^{R}\langle\chi_{3}\rangle\\
\epsilon_{23}^{L}\epsilon_{23}^{R}\langle\chi_{3}\rangle & \langle\chi_{3}\rangle
  \end{array}\right)
\end{equation}
and
\begin{equation}
m_e = \left(\begin{array}{ccc}
a_{11}^{e}\epsilon_{12}^{R}\epsilon_{23}^{R} & a_{12}^{e}\epsilon_{12}^{L}\epsilon_{23}^{R} & a_{13}^{e}\epsilon_{12}^{L}\epsilon_{23}^{L}\\
a_{21}^{e}\epsilon_{12}^{L}\epsilon_{12}^{R}\epsilon_{23}^{R} & a_{22}^{e}\epsilon_{23}^{R} & a_{23}^{e}\epsilon_{23}^{L}\\
a_{31}^{e}\epsilon_{12}^{L}\epsilon_{12}^{R}\epsilon_{23}^{L}\epsilon_{23}^{R} & a_{32}^{e}\epsilon_{23}^{L}\epsilon_{23}^{R} & a_{33}^{e}
  \end{array}\right) \, .
\end{equation}
Here we have defined $\epsilon_{ij}^{R,L}=\langle\phi_{ij}^{R,L}\rangle/\Lambda_{ij}$. The new $m_D$ is hierarchical by columns, whereas the new $M_M$ is nearly diagonal. Finally, the new charged lepton mass matrix is nearly diagonal and hierarchical. This allows one to reproduce the charged lepton masses with
\begin{equation}
\epsilon_{12}^{R}\sim\frac{m_{e}}{m_{\mu}}\simeq0.005\,,\qquad\epsilon_{23}^{R}\sim\frac{m_{\mu}}{m_{\tau}}\simeq0.06\,.
\end{equation}
If one further assumes $\epsilon_{12,23}^{L} \gtrsim 0.1$, $\langle\chi_{2}\rangle \sim 10^{13}$ GeV and $\langle\chi_{3} \rangle \sim 10^{14}$ GeV, neutrino data is also reproduced with $\mathcal{O}(1)$ coefficients. Not only that: the hierarchical column structure of the Yukawa textures in the charged lepton and neutrino sectors naturally enforces sequential dominance~\cite{King:1998jw,King:1999mb,King:2002nf,Antusch:2004gf}. This can be viewed as a specific realization of the type-I seesaw mechanism, in which the contributions of the right-handed neutrinos are hierarchical: one of them mainly generates the largest neutrino mass, while a second one predominantly accounts for the next-to-largest mass. Interestingly, this is naturally induced in our construction. As a result of this, several consequences can be derived:
\begin{itemize}
\item The neutrino mass ordering is predicted to be normal.
\item $m_1=0$ (due to the existence of only two right-handed neutrinos).
\item Naturally hierarchical charged leptons and suppressed right-handed charged lepton mixing.
\item Both sectors (charged and neutral) contribute to the lepton mixing matrix.
\item Simple analytical formulas can be derived at leading order in the sequencial dominance expansion.
\end{itemize}
In summary, the model leads to a structured and predictive description of the lepton sector, replacing anarchy with an underlying pattern.

\section{Conclusion}
\label{sec:final}

Flavor deconstruction provides a compelling framework to generate the flavor structure of the SM, with tri-hypercharge offering one of its simplest and most economical realizations. Within this approach, the quark sector is naturally well described, but the lepton sector has long posed a challenge, often requiring the assumption of anarchical structures to reproduce the observed mixing patterns. Remarkably, further decomposing tri-hypercharge changes this picture: it naturally gives rise to sequential dominance, leading to a structured and predictive pattern in the lepton sector: order emerging in place of anarchy.

\acknowledgments

The author is grateful to Mario Fern{\'a}ndez Navarro and Steve King for a very productive and enjoyable collaboration leading to the work presented here. Work supported by the Spanish grants PID2023-147306NB-I00, CNS2024-154524 and CEX2023-001292-S (MICIU/AEI/10.13039/501100011033), as well as CIPROM/2021/054 (Generalitat Valenciana).

\bibliographystyle{JHEP}
\bibliography{refs}

\end{document}